\documentclass[fleqn,twoside]{article}
\usepackage[headings]{espcrc2}

\readRCS
$Id: espcrc2.tex,v 1.2 2004/02/24 11:22:11 spepping Exp $
\usepackage{graphicx, balance}

\usepackage[figuresright]{rotating}
\usepackage{algorithmic}
\usepackage[linesnumbered, ruled, vlined]{algorithm2e}
\usepackage[caption=false]{subfig}

\newcommand{\AmS}{{\protect\the\textfont2
  A\kern-.1667em\lower.5ex\hbox{M}\kern-.125emS}}
\usepackage{amsmath}
\title{\textbf{Secret Image Sharing Based Cheque Truncation System with Cheating Detection}}

\author{Sreela.S.R\address[DCSE]{Department of Computer Science, Cochin University of Science and Technology \\},
G. Santhosh Kumar\address{Department of Computer Science, Cochin University of Science and Technology \\},
Binu.V.P\address{Department of Computer Applications, Cochin University of Science and Technology}}

\runtitle{Secret Image Sharing Based CTS with Cheating Detection}
\runauthor{Sreela S R, et al.,}

\begin{document}
\begin{abstract}

Cheque Truncation System(CTS) is an automatic cheque clearance system implemented by RBI.CTS  uses cheque image, instead of the physical cheque itself, 
for cheque clearance thus reducing the turn around time drastically. This approach holds back the physical movement of cheque from
presenting bank to the drawee bank. In CTS, digital image of the cheque is protected using standard public key and
symmetric key encryptions like RSA, triple DES etc. This involves a lot of computation overhead and key management.
The security also depends on the hard mathematical problem and is only computationally secure.Information
theoretically secure, secret image sharing techniques can be used in the CTS  for the secure and efficient processing
of cheque image .In this paper, we propose two simple and efficient secret image sharing schemes and a Cheque Truncation System based on these algorithms . 
In the proposed scheme,the presenting bank is acting as the dealer and the participants are the customer, and the drawee bank.The dealer should generate the shares 
of cheque and distributes it to customer and drawee bank.The validity of the shares are important during the reconstruction process.
The proposed scheme also suggests a method for cheating detection which identify any invalid shares submitted by the customers, using the hashing technique.
The experimental results shows that the proposed scheme is efficient and secure compared with the existing scheme.
\\\\
{\bf Keywords :} Cheque Truncation System,Secret image sharing,PKI,Pixel expansion,Visual cryptography.
\end{abstract}

\maketitle

\section{INTRODUCTION}
Cheques represent a significant segment of payment instruments in India. Cheque Truncation System (CTS) or ICS(Image Based Clearing System) in India is a project undertaken by Reserve Bank of India ( RBI) for faster clearing of cheques. CTS is basically an online image-based cheque clearing system where cheque images and Magnetic Ink Character Recognition (MICR) data are captured at the collecting bank branch and transmitted electronically.Manual clearing of cheque needs human intervention and is a time consuming task.Cheque truncation \cite{rbi} involves stopping the flow of the physical cheques issued by a drawer to the drawee branch. An electronic image of the cheque is sent to the drawee branch along with the relevant information like the MICR fields, date of presentation, presenting banks etc.
The point of truncation is left to the discretion of the presenting bank. Thus Cheque truncation, would eliminate the need to move the physical instruments across branches and hence result in effective reduction in the time required for payment of cheques, the associated cost of transit and delays in processing, etc.
This will  speed up the process of collection or realization of cheques and thus reduce the turn around time.

The system offers following benefits to the bank and customers. Banks can expect multiple benefits through the implementation of CTS, like faster clearing cycle,better reconciliation/verification process. Besides, it reduces operational risk by securing the transmission route.Reduction of manual tasks leads to reduction of errors. Customer satisfaction will be enhanced, due to the reduced turn around time (TAT). Real-time tracking and visibility of the cheques, less fraudulent cases with secured transfer of images to the RBI are other possible benefits that banks may derive from this solution \cite{cts}.
For Customers CTS / ICS substantially reduces the time taken to clear the cheques as well  increases 
operational efficiency by cutting down on overheads involved in the physical cheque clearing process. 
In addition, it also offers better reconciliation and fraud prevention. \\
The use of the Public Key Infrastructure (PKI) ensures data authenticity, integrity and non-repudiation, adding strength to the entire system. The presenting bank is required to affix digital signature on the images and data from the point of truncation itself. The image and 
data are secured using the PKI through out the entire cycle covering capture system,the presenting bank, the clearing house and the drawee bank.This system needs a lot of computation and overhead in key management is high.In this paper a secret image sharing \cite{sisvc} based scheme is proposed.Two efficient schemes are proposed which are computationally secure and avoids the overhead in key management.A cheating
detection scheme is also proposed which avoids the use of invalid shares during the reconstruction.\\
In the rest of the paper, section 2 describes the CTS Architecture. Section 3 describes the related work.Proposed system and algorithms are explained in section 4.  Experimental results are discussed in section 5 and the  conclusions are drawn in section 6.
\section{CTS ARCHITECTURE}
\begin{figure}[Ht]

\centering
\includegraphics[width=7.5cm,height=5cm]{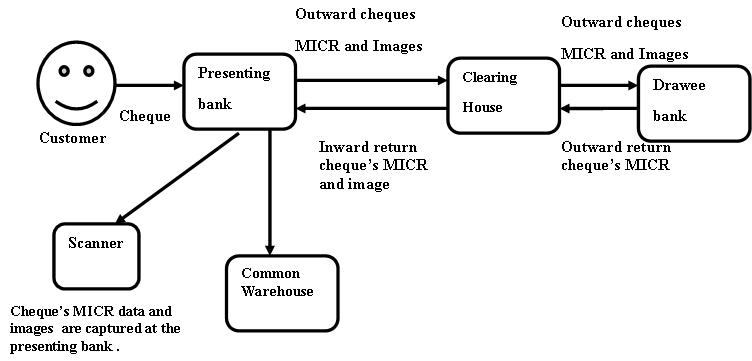}
\caption{CTS Architecture}
\label{fig:cts}
\end{figure}

The process flow of CTS is explained below. In CTS, the presenting bank (or its branch) captures the data on the MICR band and the images of a cheque using their Capture System comprising of a scanner, core banking or other application. Images and data should meet the specifications and standards prescribed for data and images. The architecture of CTS is explained in figure \ref{fig:cts}. \\
To ensure security, end-to-end Public Key Infrastructure (PKI) has been implemented in CTS for protecting data and image. The presenting bank sends the data and captured images duly signed and encrypted to the Clearing House (the central processing location) for onward transmission to the paying bank (destination or drawee bank). For the purpose of participation the presenting and drawee banks are provided with an interface / gateway called the Clearing House Interface (CHI) that enables them to connect and transmit data and images in a secure and safe manner to the Clearing House (CH).
The CTS uses public key infrastructure(PKI) like digital signature and encryption for protecting cheque images and data. The standards defined for PKI are hash algorithm SHA-1, padding algorithm,RSA asymmetric encryption with 1024 bit key length, Triple DES (3DES, TDES) symmetric encryption with 168 bit key length, and Certificates in x.509v3 format.Cheque image is protected using encryption techniques. These techniques need a lot of computation and usage of keys.\\ 

\section{RELATED WORK}
CTS system is implemented by RBI to reduce the complexity of cheque processing. CTS system is implemented in India in 2010. Grid based CTS is implemented in Chennai, Delhi, Kolkata etc. The different security schemes are applied in cheque.Pasupathinathan, Vijayakrishnan, Josef Pieprzyk, and Huaxiong Wang \cite{ecs} describes  privacy enhanced electonic cheque system in 2005. In 2011, Rigel Gjomemo,Ha.z.Malik, Nilesh Sumb,V.N.
Venkatakrishnan and Rashid Ansari  \cite{forgery} explains the digital cheque Forgery attack on CTS.  Kota, Saranya, and Rajarshi Pal\cite{watermarking} explains the method for detecting tampered cheque images in CTS Using Difference Expansion Based Watermarking in 2014.\\
The secret image sharing schemes are based on visual cryptography, number theory \cite{lossless}, information hiding theory, error diffusion technique, boolean operation etc.  In Yan, Xuehu scheme \cite{yan2013new}, secret image sharing is based on information hiding theory. The important technique used in this scheme are MLE and LSBM. But this scheme is applicable only to the binary images. Chen and Chang \cite{chen2007secret} use quadratic residue technique for secret image sharing. They proposed a (2, 2) scheme which is lossy and the lossless scheme having the share size larger than the secret. The computations involved is also more. Chen, Wei-Kuei, and Hao-Kuan Tso \cite{visualchen2013} introduced a secret image sharing scheme for protecting medical images using Hill cipher method. Thein-Lin’s Scheme enhanced Shamir's secret sharing scheme  \cite{shamir1979share}(Lagrange interpolating polynomial) for protecting digital images.\\
Table1 explains a comparative study on different secret image sharing schemes.

\begin{table}[H]
\caption{\bf Comparison of different schemes}

\begin{tabular}{|l|l|l|l|l|l|}

\hline Scheme &\parbox[t]{1.2cm}{(k,n)\\threshold} & \parbox[t]{2cm}{Recovering\\ Measure} & \parbox[t]{1cm}{Loss-\\less} & \parbox[t]{1.5cm}{ Pixel\\ Expansion} & \parbox[t]{1cm}{size of\\ share}\\[4pt] \hline
VCS \cite{visualcryptography} & (k,n) & Stacking & No & Yes & Increases\\ \hline
\parbox[t]{1cm}{Extended \\ VCS \cite{extended}} & (k,n)  & Stacking & No & Yes & Increases \\ \hline
(k, n2) \cite{yan2013new} & (k, n2)  & \parbox[t]{2cm} {Mod and\\ Boolean /\\ addition and \\ comparison} & Yes & No & \parbox[t]{1cm}{same as \\ original image}\\ \hline

(2,3) \cite{lossless} & (2,3)  & \parbox[t]{1cm}{Mod and\\ Multiplication} & Yes & No & \parbox[t]{1cm}{same as \\ original image}\\ \hline
\parbox[t]{1cm}{Thein-Lin} \cite{sis} & (r , n)  & \parbox[t]{1cm}{Shamir’s SSS \\(Lagrange \\Interpolating\\ polynomial)} & No & No & \parbox[t]{1cm}{Reduces\\ by half} \\ \hline
\parbox[t]{1cm}{Boolean \\ VSS} \cite{boolean} & (r , n) &\parbox[t]{2cm}{Boolean\\operations} & Yes  & No & Increases \\

\hline
\end{tabular}
\end{table}
\section{PROPOSED SYSTEM}
\begin{figure}[t]
\centering

\includegraphics[width=7.5cm,height=5.2cm]{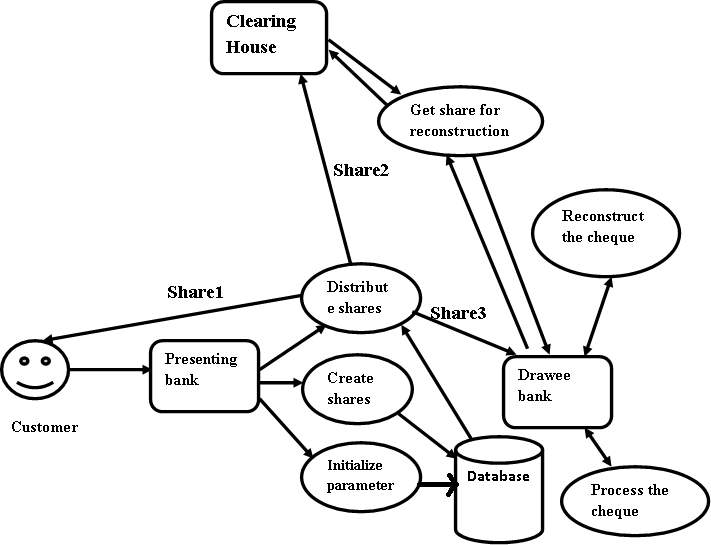}
\caption{System Architecture}
\label{fig:ctssisarch}
\end{figure}

\begin{figure}[Ht]
\centering
\includegraphics[width=7.5cm,height=5.2cm]{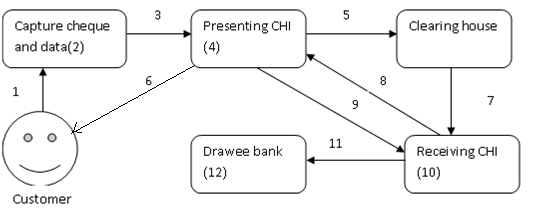}
\caption{System Architecture}
\label{model}
\end{figure}

The system architecture describes how secret image sharing scheme happening in the CTS. In this architecture, the dealer should be the presenting bank. The participants are customer, clearing house (CH), and drawee bank. Figure \ref{fig:ctssisarch} explains the system architecture. \\
In order to reduce the computation and usage of keys, cheque image can be protected using secret image sharing.
In this paper, two secret image sharing methods are proposed  for protecting cheque images.If any one of the participants do malpractice on the shares, then cheating occurs. Cheating detection is implemented in this paper.\\In secret image sharing technique, a secret image is distributed to some of the participants through splitting the image into different pieces called shares and recover the secret image by collecting the sufficient number of shares from authorized participants. This field of cryptography is called visual cryptography or visual secret sharing \cite{visualcryptography}. If any one of the participant do malpractice on their shares, cheating detection methods can be used. It consists of three phases: share generation phase, distribution phase and reconstruction phase.In the share generation phase, the digital image is split into different pieces called shares. In the distribution phase, the shares are distributed to authorized participants and in the last phase, the image is reconstructed using sufficient number of shares from authorized participants.In a secret image sharing scheme, there is a secret image S to be shared among a set of participants. The secret is known to a special person called dealer. The dealer generates and distributes partial information called shares to the participants. \\
(2,3) scheme is required for implementing security in CTS. Presenting bank generates the shares and distributed to the clearing house, drawee bank and to the customer. Customer should use the share to get the information of processing of cheque through online. Drawee bank should reconstruct the cheque image using the share from the CH and its own share. Drawee bank can’t reconstruct the cheque image using its own share. To implement security in CTS, xor scheme and partition scheme can be used.\\
The important steps involved in the proposed CTS using secret image sharing are as follows:
\begin{enumerate}
\item Customer submits the cheque to the presenting bank.
\item Capture image of cheque and data using capture system
\item Send the data and image to the presenting CHI.
\item Presenting CHI provide security to the cheque image using (3,2) secret image sharing scheme.
\item Send first share of the cheque image(SC1) and data to the clearing house through the CHI.
\item Send second share of the cheque image (SC2) to the customer if customer submits cheque through online and this share is used for authentication for viewing the details of cheque processing.
\item The clearing House send data and one share of the cheque image(SC1) to the drawee bank through receiving CHI.
\item The drawee bank request another share of the cheque image from the presenting bank through receiving CHI.
\item The presenting bank submit third share (SC3) to the drawee bank.
\item The receiving CHI reconstructs the cheque image using shares SC1 and SC3.
\item Send data to the drawee bank for processing cheque.
\item Bank process the cheque using image processing algorithm.

\end{enumerate}
In figure \ref{model}, the numbers represent the above steps.

\subsection{XOR scheme}
XOR scheme is a (2,3) scheme. In this scheme, three shares are created and the original image is reconstructed using at least two shares. The image is not reconstructed using only one share.The share images are created by dividing pixel into four bits.In this scheme, the share image pixel is 4 bits. Share generation algorithm is explained in Algorithm \ref{algo:shrgen}. Recovery algorithm is explained in Algorithm \ref{algo:rec}.The original secret image is reconstucted by using any of the two shares from three shares. \\
Consider an image matrix is
$$\begin{bmatrix}
157&160&190&130\\
89&255&224&192 \\
10&220&255&224 \\
64&128&192& 255 
 \end{bmatrix}$$
Consider the secret image pixel is 190. Its binary representation is 10111110. The share1 pixel(sc1=6(0110)) is created using even bits.The share2 pixel(sc2=15 (1111)) is created using odd bits. The share3 pixel(sc3=9(1001)) is created by xor of s1 s2. The xor scheme is applied on the above image S. The three shares obtained SC1, SC2 and SC3 are as follows:\\
SC1=$$\begin{bmatrix}
7&0&6&0\\
13&15&8&8 \\
0&14&15&8 \\
8&0&8& 15 
 \end{bmatrix}$$\\
 SC2=$$\begin{bmatrix}
 10&12&15&9\\
 2&15&12&8 \\
 3&10&15&12 \\
 0&8&8& 15 
  \end{bmatrix}$$\\
   SC3=$$\begin{bmatrix}
   13&12&9&9\\
   15&0&4&0 \\
   3&4&0&4 \\
   8&8&0&0 
    \end{bmatrix}$$\\
\begin{algorithm}
\begin{scriptsize}
 Input: M X N Secret grayscale image S\\
 \indent Output: Share images SC1, SC2, SC3 \\
begin\\
\begin{enumerate}

\item For each pixel $(i,j)\varepsilon\{(i,j)| 1\leq i \leq M, 1\leq j \leq N\} $ repeat steps 2- 4
\item Pixelvalue, pv= S(i,j) which is the binary array containing the pixel intensity binary representation.
\item Create share1 SC1(i, j) pixel using even bits of  S(i,j) pixel 
    \[ SC1(i,j)= \sum_{k=0}^{3}(pv(2k)\times 2^k)\]     
\item Create share2 SC2(i,j) pixel using odd bits of S(i,j) pixel  
    \[ SC2(i,j)= \sum_{k=0}^{3}(pv(2k+1)\times 2^k)\] 
\item Create share3 SC3(i,j) pixel by xor ing  SC1(i, j) and SC2(i, j) 
    \[SC3(i,j,k)= SC1(i,j)\oplus SC2(i,j) \]
\item Output shares SC1, SC2, and SC3.
\end{enumerate}

${\bf}$end\\

\caption{Algorithm: Share generation } 

\label{algo:shrgen}
\end{scriptsize}
\end{algorithm}
\begin{algorithm}
\begin{scriptsize}
 Input: Share images SC1, SC2, SC3 \\
 \indent Output: Reconstructed Secret image S \\
begin\\
The secret image can be reconstructed from shadow images SC1, SC2  \\
\begin{enumerate}
\item For each position, $(i,j)\varepsilon\{(i,j)| 1\leq i \leq M, 1\leq j \leq N\}$ repeat  step2. \\
\item S( i , j)  is obtained by intermixing bits of SC1(i, j) and SC2(i, j) in even and odd positions respectively.     \\
\item Output image S.\\
\end{enumerate}
The secret image can be reconstructed from shadow images SC1, and SC3   \\
\begin{enumerate}
\item For each position,$(i,j)\varepsilon\{(i,j)| 1\leq i \leq M, 1\leq j \leq N\}$ repeat  step 2-3.\\
\item $ b= SC1(i, j) \oplus  SC3(i, j)  $\\
\item S(i, j)  is obtained by intermixing bits of SC1(i,j) and b in even and odd positions respectively.\\
\item Output image S.\\
\end{enumerate}
The secret image can be reconstructed from shadow images SC2, and SC3.\\
\begin{enumerate}
\item For each position, $(i,j)\varepsilon\{(i,j)| 1\leq i \leq M, 1\leq j \leq N\}$  repeat  step 2-3
\item $ b= SC2(i, j)\oplus  SC3(i, j) $
\item S(i, j)  is obtained by intermixing bits of b and SC2(i,j) in even and odd positions respectively.
\item Output image S.
\end{enumerate}
${\bf}$end\\
\caption{Recovery algorithm } 

\label{algo:rec}
\end{scriptsize}
\end{algorithm}
In this scheme, the size of the share is half of the size of the original image. The number of bits for representing a pixel in share is 4 bits.If the $M \times N$ secret gray scale image has a size of $8\times M \times N$ bits, then the size of the share is only $4 \times M \times N$ bits. So the storage space of the share is reduced. The quality of the reconstructed image is same as the original image. In this scheme, there is no pixel expansion on reconstructed image. It is a lossless scheme.
\subsection{Partition Scheme }
\begin{algorithm}
\begin{scriptsize}
 Input: M X N Secret grayscale image S\\
 \indent Output: Share images SC1, SC2, SC3 \\
begin
 \begin{enumerate}
\item Let s be the pixel of the secret image(S) and r be a random number in {0-255}. 
\item s is divided into s1 and s2 and r into r1 and r2.    
\item Share1 pixel is created by combining  $ s2\oplus  r2 $ and r1  
\item Share2 pixel is created by combining    
    $ s1\oplus  r1 $ and r2
\item Share3 pixel is created by combining
    $ s2\oplus  r1$ and $ s1 \oplus r2 $ 
\item  repeat step 1-5 until all pixels of image are processed. 
\item Output three shares share1(SC1), share2 (SC2), share3(SC3). 
\end{enumerate}
\caption{Algorithm: Share generation}
\label{algo:shrgen1}
\end{scriptsize}
\end{algorithm}
Partition scheme is a (2,3) scheme.It uses boolean xor operations. This method uses random number for creating shares. The share generation algorithm is explained in Algorithm \ref{algo:shrgen1} . Algorithms \ref{algo:ren11}, \ref{algo:ren12}, \ref{algo:ren13} describe the reconstruction of image.

\begin{algorithm}
\begin{scriptsize}
Input: Share images SC1, SC2 \\
 \indent Output: Reconstructed Secret image S \\
begin\\
Original image is reconstructed from share1 and share2 by applying following steps.
    \begin{enumerate}
        \item  The share1(sc1) pixel is divided into two equal parts sc11 and sc12.
        \item  The share2(sc2) pixel is divided into two equal parts sc21 and sc22. 
        \item The second part of the original image pixel (s2) is reconstructed by XOR-ing first part of the share1 pixel(sc11) and second part of the share2 pixel(sc22).
        \[ s2= sc11 \oplus sc22 \]
        \item The first part of the original image pixel(s1) is reconstructed by XOR-ing second part of the   the share1 pixel(sc12) and first part of the share2 pixel(sc21).
        \[ s1= sc12 \oplus sc21 \]
        \item The original image pixel(s) is obtained by combining s1 and s2. 
        \[s=s1.s2\] 
        \item repeat the above steps until all pixels are processed. 
        \item Output image S        
        \end{enumerate} 
       
${\bf}$end\\
\caption{Algorithm: Reconstruction using share1 and share2}
\label{algo:ren11}
\end{scriptsize}
\end{algorithm}


\begin{algorithm}
\begin{scriptsize}
Input: Share images SC1, SC3 \\
 \indent Output: Reconstructed Secret image S \\
begin\\
        Original image is reconstructed from share1 and share3 by applying following steps. 
        \begin{enumerate}
                \item The share1 pixel is divided into two equal parts sc11 and sc12.
                \item The share3 pixel is divided into two equal parts sc31 and sc32.
                \item The second part of the original image pixel(s2) is obtained by XOR-ing second part of the share1 pixel(sc12) and first part of the share3 pixel(sc31).
                \[ s2= sc12 \oplus sc31 \]
                \item  $ b = s1 \oplus s2$ is obtained by XOR-ing first part of the share1 pixel(sc11) and second part of the share3 pixel(sc32).
                \[   b = sc11 \oplus sc32\]
                \item The first part of the original image pixel(s1) is obtained by
                $ s1=b \oplus s2 $
                \item Secret image pixel(s) is reconstructed by combining s1 and s2
                \[s=s1.s2\]
                \item Repeat above steps until all pixels are processed.
                \item Output image S
                \end{enumerate}  
    
${\bf}$end\\
\caption{Algorithm: Reconstruction using share1 and share3}
\label{algo:ren12}
\end{scriptsize}
\end{algorithm}
\begin{algorithm}
\begin{scriptsize}
Input: Share images SC2, SC3 \\
 \indent Output: Reconstructed Secret image S \\
begin\\  
                Original image is reconstructed from share2 and share3 by applying following steps.
                \begin{enumerate}
                        \item The share2 pixel is divided into two equal parts sc21 and sc22.
                        \item The share3 pixel is divided into two equal parts sc31 and sc32.
                        \item The first part of the original image pixel(s1) is obtained by XOR-ing second part of the share2 pixel(sc22) and second part of the share3 pixel(sc32).
                        \[ s1= sc22 \oplus sc32 \]
                        \item  $b = s1 \oplus s2$ is obtained by  XOR-ing first part of the share2 pixel(sc21) and first part of the share3 pixel(sc31).
                        \[ b = sc21 \oplus sc31 \]
                        \item The second part of the original image is obtained by $ s2=b \oplus s1 $
                        \item Secret image pixel(s) is reconstructed by combining s1 and s2.
                       \[s=s1.s2\]
                        \item Repeat above steps until all pixels are processed.
                         \item Output image S
                        \end{enumerate}
    
${\bf}$end\\
\caption{Algorithm: Reconstruction using share2 and share3}
\label{algo:ren13}
\end{scriptsize}
\end{algorithm}
\subsection{Cheating detection scheme using Hash function}
A threshold scheme for secret sharing
can protect a secret with high reliability and flexibility.These advantages can be achieved only
when all the participants are honest, i.e. all the
participants willing to pool their shadows shall
always present the true ones. Cheating detection is
an important issue in the secret sharing scheme.
However, cheater identification is more effective
than cheating detection in realistic applications. If
some dishonest participants exist, the other honest
participants will obtain a false secret, while the
cheaters may individually obtain the true one.By applying a one-way hashing function along with the use of arithmetic coding, the proposed method can be
used to deterministically detect cheating and
identify the cheaters, no matter how many
cheaters are involved in the secret reconstruction.\\
Two important theorems used in cheating detection using hash function are as follows.
Let $a_i$ be the random shares of the secret data and $p$ be the randomly generated prime number.\\
Theorem 1 \cite{cheat1995}:
Let $T=\sum_{i=1}^{n} a_ip^{i-1},$ where $0\leq a_i < p $. Then 
\begin{equation}
\lfloor \frac{T}{p^{j-1}}\rfloor(mod \quad p)=a_j
\end{equation}\\
 Extended from Theorem 1, we have the following result.\\
 Theorem 2 \cite{cheat1995}:
  Let $T=\sum_{i=1}^{n}a_ip^{2(i-1))}+ \sum_{i=1}^{n-1}cp^{2i-1}$, where $-p<a_i<p$ and $1\leq c < p$. Then
 \begin{equation}
  \lfloor \frac{T}{p^{2(j-1)}}\rfloor (mod \quad p)= a_j(mod\quad p)
 \end{equation}
 Combining this result with secret image sharing scheme, the following method is used for cheating detection and cheater identification.Algorithm for cheating detection and cheater identification is explained in Algorithm \ref{algo:sishash}.\\
 \begin{algorithm}
 \begin{scriptsize}
 Dealer generates the shares for cheque image using secret image sharing algorithm.\\
 He generates public parameters T and p in the following steps.\\

 Choose a one-way function $h ( . )$ and a prime
 number p such that $h( .) < p$.
 Generates hash value of image using hash function.
 \\
  Compute $T=\sum_{i=1}^{n}h(s_i)p^{(2(i-1))}+ \sum_{i=1}^{n-1}cp^{2i-1}$ where $c$ is a positive constant randomly chosen over $GF(p)$\\
  Publish T and p.\\
   Dealer distributes shadow $SC_i$ to participants $U_i$ . for $i = 1, 2, . . ., n$.\\
   In the receiver side, cheating detection and cheater identification can easily be
   achieved by applying the following procedure.\\
   Participants $U_j\epsilon G$ present their possessed shadows $SC_j'$  and compute
   $T'=\sum_{U_j\epsilon G}h(SC_j')p^{(2(i-1))}$\\
   For each $U_j\epsilon G$, check
   $\lfloor \frac{T-T'}{p^(2(j-1))}\rfloor (mod\quad p) \stackrel{?}{=}0$\\
   If the equation holds, participant$U_j$ is honest; otherwise, $U_j$ is a cheater.
   
 \caption{Cheating detection and cheater identification using hash function }
 \label{algo:sishash}
 \end{scriptsize}
 \end{algorithm}
\\
The hash value of the image is generated using content of the image. The hash value of the image is also generated using feature vector of the image.In the secret image sharing, any simple change in the shares is treated as a cheating. Any mild change in the image is reflected in the hash value of image using content of image rather than using feature vector of image.So we use the hash generation method using content of the image.
\subsection{Cheque processing}
Cheque processing is implemented in Drawee bank. In our work, the courtesy amount region and account number field is recognized.The important steps associated with cheque processing are as follows:
\begin{enumerate}
\item	Load cheque in grayscale.
\item	Find courtesy amount region in cheque using cheque template method .
\item	Find account number region in cheque using cheque template method.
\item	Segment digits in courtesy amount and resize each digit having a size of $28 \times 28$.
\item	Apply digit recognition method for recognizing digit in courtesy amount.
\item	Combine each digit and generate courtesy amount.
\item	Segment digits in account region  and resize each digit having a size of $28 \times 28$.
\item	Apply digit recognition method for recognizing digit in account number.
\item	Combine each digit and generate account number.
\item	Process the amount from the account and deduct the amount from the account.
\item	Send the information to the presenting bank through clearing house.
\item	At last customer gets the amount from the presenting bank. 
\end{enumerate}

\subsubsection{Digit recognizer}
In our work, digit recognition is done using $K$ Nearest neighbour classification technique \cite{knn}.
The isolated components after segmentation are fed into a digit recognizer.The accuracy in recognizing constituent digits plays a big role in the recognition accuracy of the handwritten courtesy amount numeral string. After successful segmentation
of individual digits from the numeral string, they have to be correctly recognized to get the value of the cheque.In this, there is two steps: training phase and testing phase.In training phase, hand written images are trained. In the testing phase,the following steps need to be carried out.
\begin{itemize}
\item The digit in the image is centered.
\item Convert two dimensional array to one dimensional array using reshape operation.
\item Apply the one dimensional array to the KNN classifier.
\item The digit is recognized as output.
\end{itemize}
\section{EXPERIMENTAL RESULTS}

\begin{figure}[t]
\centering
\subfloat[Secret image]
     {
         \includegraphics[width=0.15\textwidth]{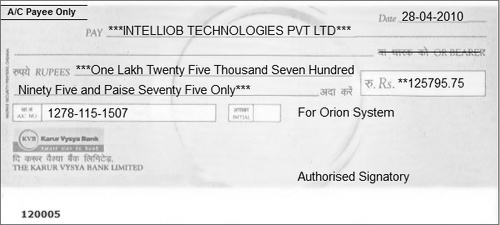}
         \label{fig:first_sub}
     }
     \\
     \subfloat[Share1]
     {
         \includegraphics[width=0.15\textwidth]{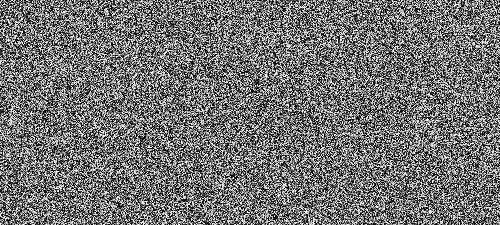}
         \label{fig:second_sub}
     }
     \subfloat[Share2]
     {
         \includegraphics[width=0.15\textwidth]{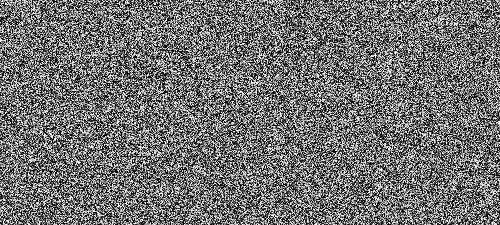}
         \label{fig:third_sub}
     }
     \subfloat[Share3]
          {
         \includegraphics[width=0.15\textwidth]{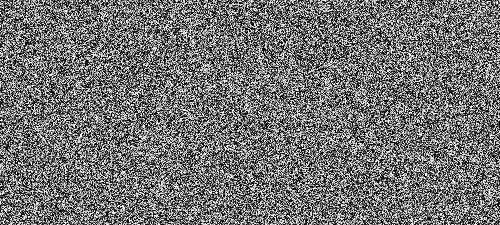}
              \label{fig:third_sub}
          } \\
     \subfloat[Reconstructed image]
               {
              \includegraphics[width=0.15\textwidth]{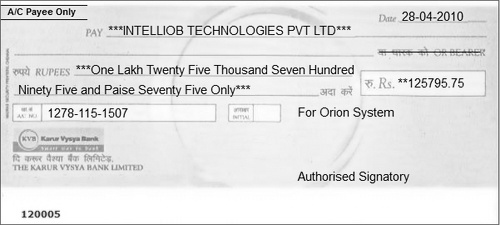}
                   \label{fig:third_sub}
               }     
     \caption{Result of partition scheme}
     \label{fig:sample_subfigures}
\end{figure}

The algorithms are implemented in Java.
The experimental result obtained for partition scheme using the $500 \times 225$ gray scale cheque image is shown in figure\ref{fig:sample_subfigures}. The reconstructed image has the same quality as original image.This algorithm is also useful for color images.If this algorithm is applied in color images, the algorithm is applied on each channel (Red, Blue, Green) separately.The bit depth of the share of color image is 12 bits.So this scheme is enhanced for color images also. The comparison of above schemes are described in table2.\\
\clearpage
\begin{table}[t]
\caption{\bf Property comparison of proposed schemes}

\begin{tabular}{|l|l|l|l|l|l|}

\hline Scheme  & \parbox[t]{1.5cm}{Recovering\\ Measure} & \parbox[t]{1cm}{Loss-\\less} & \parbox[t]{1cm}{ Pixel\\ Expan\\sion} & \parbox[t]{1cm}{size of\\ share}\\[4pt] \hline

xor & Boolean  & Yes  & No & \parbox[t]{1cm}{half of \\the image size} \\[4pt] \hline
Partition & Boolean  & Yes  & No & \parbox[t]{1cm}{same as\\ cheque image}\\[4 pt]

\hline
\end{tabular}
\end{table}
The mean square error(MSE) is used to measure the mean square error between original(I) and recovered image(I') and is calculated by using the equation 
\[MSE= \frac{1}{MN} \sum_{i=1,M}\sum_{j=1,N}(I'(i,j)-I(i,j))^2\]
The MSE between original and recovered image is 0.\\

\begin{figure}[h!]

 \includegraphics[width=0.5\textwidth]{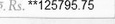}
        
         \caption{Courtesy amount}
          \label{fig:courtesyamt}
\end{figure}

In the cheating detection phase, the hash value of the share images are calculated in the sender side. The value of T is 4.59080713E8. In the receiver side, the hash value is not computed. The value of T' is computed in the receiver side. If the remainder is zero, the cheating does not occur in the shares of the cheque image. If the cheating does not occur in the shares, the cheque image is reconstructed from the shares.Otherwise, the drawee bank request for the correct shares from the participants. \\
In Cheque processing, the courtesy amount and account number region are recognised using image processing technique.The courtesy amount region in cheque image is shown in fig. \ref{fig:courtesyamt}

\begin{figure}[h!]

 \includegraphics[width=0.5\textwidth]{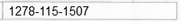}
        
         \caption{Account number}
          \label{fig:accountno}
\end{figure}

The account number region in cheque image is shown in fig. \ref{fig:accountno}
Each digit in courtesy amount and account number are segmented and applied to the digit recognizer.
For digit recognition using KNN, the standard dataset MNIST handwritten digit image is used as training set. MNIST dataset contains 60000 image for training purpose. The KNN classifier give correct result for MNIST testing images.  Some misclassification occured for courtesy amount recognition. The courtesy amount recognised in fig. \ref{fig:courtesyamt} is 125795.75. The account number reconised in fig. \ref{fig:accountno} is 12781151507.The account number and amount is fed to the core banking software and do the transaction operations in software. Drawee bank returns the transaction details or error message to the presenting bank through Clearing House.
\section {CONCLUSIONS}
Cheque Truncation system accelerates the
process of collection of cheques resulting in
better service to customers, reduces the scope
for clearing-related frauds or loss of instru-
ments in transit, lowers the cost of collection
of cheques, and removes reconciliation-related
and logistics-related problems, thus benefit
the system as a whole.In this paper, two secret
image sharing schemes are proposed for provid-
ing security to the cheque image in the CTS.
The proposed XOR scheme is simple and effecient but it is not ideal.It can be used in low storage device where memory is a contsraint.The share size is only half of the original image and it is a lossless scheme. Partition scheme have the properties such as no pixel expansion and lossless
scheme. The scheme is also ideal.\\

The experimental result shows that the proposed system provides better security and efficiency in Cheque Truncation System.The operations invloved are simple XOR and it also avoids the complicated encryption decryption operations which are time consuming.The secret image sharing  scheme doesn't need any key management
and authentication of the shares are done with simple hash function.The shares are also verified  with the help of public parameters.We are looking forward
for  improved cheque processing  using advanced image
processing technique which helps in automatic cheque processing.The operational effeciency, speed
accuracy, security and authentication are the major design objectives.	

\noindent{\includegraphics[width=1in,height=1.7in,clip,keepaspectratio]{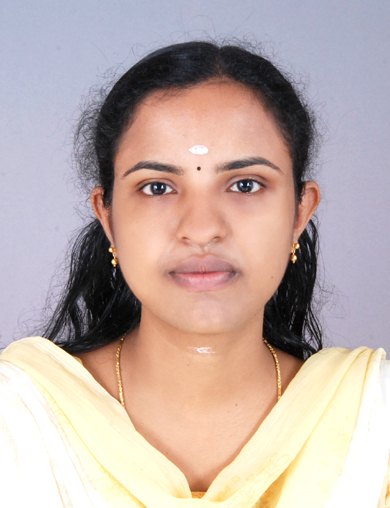}}
\begin{minipage}[b][1in][c]{1.8in}
{\centering{\bf {S R Sreela}} is a Research Scholar in the Department of Computer Science, Cochin University of Science and Technology(CUSAT).She holds a Bachelor Degree in Information Technology and Masters Degree in Computer } \\ \\
\end{minipage}
 and Information Science.Her research area includes image processing, secret sharing and security. \\ \\
\noindent{\includegraphics[width=1in,height=1.7in,clip,keepaspectratio]{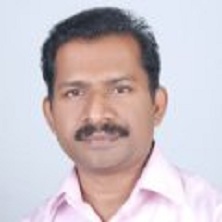}}
\begin{minipage}[b][1.4in][c]{1.8in} 
{\centering{ \bf{Dr.G.Santhosh Kumar}} received his MTech Degree in Computer and Information Science from  CUSAT, in 1999 and PhD in Wireless Sensor Network from Cochin University of Science and Technology.} 
\end{minipage}
Currently he is working as an Associate Professor in CUSAT. He had more than 15 years of teaching experience. His research interest includes Wireless Networks, Mobile Communications and Software Architecture.\\ \\
\noindent{\includegraphics[width=1in,height=1.7in,clip,keepaspectratio]{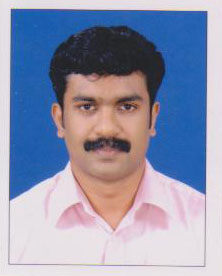}}
\begin{minipage}[b][1in][c]{1.8in}
{\centering{\bf {V P Binu}} is a Research Scholar in the Department of Computer Applications, Cochin University of Science and Technology(CUSAT).He holds a Bachelor Degree in Computer Science and Engineering and Masters Degree  } \\ \\
\end{minipage}
in Computer and Information Science.His research area includes cryptography, secret sharing and security. \\ \\

\end{document}